\documentclass[traditabstract]{aa}
\usepackage[utf8]{inputenc}
\usepackage{amsmath}
\usepackage{amssymb}
\usepackage{microtype}
\usepackage{natbib}
\usepackage{graphicx}
\usepackage{txfonts}
\usepackage{color}
\usepackage{multirow}
\usepackage{xspace}
\bibpunct{(}{)}{;}{a}{}{,} 

\newcommand{\src}{\object{1A~0535$+$262}\xspace}

\title{\emph{XMM-Newton} observations of \src in quiescence.}
\author{V.\,Doroshenko\inst{1}, A.\,Santangelo\inst{1}, R.\,Doroshenko\inst{1}, I.\,Caballero\inst{2}, S.\,Tsygankov\inst{3,4,5}, R.\,Rothschild\inst{6}}	
\institute{Institut für Astronomie und Astrophysik, Sand 1, 72076 Tübingen, Germany\and
AIM-CEA Saclay, Paris, France\and
Finnish Centre for Astronomy with ESO (FINCA), University of Turku,
V\"ais\"al\"antie 20, FI-21500 Piikki\"o, Finland\and
Astronomy Division, Department of Physics, PO Box 3000, FI-90014 University of Oulu, Finland\and
Space Research Institute of the Russian Academy of Sciences,
Profsoyuznaya Str. 84/32, Moscow 117997, Russia\and
University of California, San Diego, Center for Astrophysics and Space Sciences, 9500 Gilman Dr., La Jolla, CA 92093-0424, USA
}

\begin{document}

\bibliographystyle{aa}

\abstract{Accretion onto magnetized neutron stars is expected to be
centrifugally inhibited at low accretion rates. Several sources, however, are
known to pulsate in quiescence at luminosities below the theoretical limit
predicted for the onset of the centrifugal barrier. The source \src is one of
them. Here we present the results of an analysis of a $\sim50$\,ks long
\emph{XMM-Newton} observation of \src in quiescence. At the time of the
observation, the neutron star was close to apastron, and the source had
remained quiet for two orbital cycles. In spite of this, we detected a pulsed
X-ray flux of $\sim3\times10^{-11}{\rm erg\,cm}^{-2}{\rm\,s}^{-1}$. Several
observed properties, including the power spectrum, remained similar to those
observed in the outbursts. Particularly, we have found that the frequency of
the break detected in the quiescent noise power spectrum follows the same
correlation with flux observed when the source is in outburst. This correlation
has been associated with the truncation of the accretion disk at the
magnetosphere boundary. We argue that our result, along with other arguments
previously reported in the literature, suggests that the accretion in
quiescence also proceeds from an accretion disk around the neutron star. The
proposed scenario consistently explains the energy of the cyclotron line
observed in \src, and the timing properties of the source including the spin
frequency evolution within and between the outbursts, and the frequency of the
break in power spectrum.}

\authorrunning{V. Doroshenko et al.}
\maketitle

\section{Introduction} The source\src is one of the best-studied high mass
X-ray binaries (HMXBs) of the Galaxy. The system hosts a \emph{Be} star
\citep{Giangrande80} and a neutron star in an eccentric orbit with $e\sim0.47$
and orbital period of $\sim111$\,d \citep{Finger96}. The neutron star is an
accreting pulsar with spin period of $\sim103.3$s \citep{Rosenberg75}, which
exhibits outbursts associated with passage through the circumstellar disk of
the primary in the vicinity of periastron. The X-ray luminosity reaches
$10^{37-38}{\rm erg\,s}^{-1}$ during the so-called normal and giant outbursts
(assuming a distance of 2\,kpc; \citealp{Steele98}). The magnetic field of the
neutron star has been estimated to be $B\sim5\times10^{12}$\,G from the
centroid energy of the so-called cyclotron resonance scattering feature (CRSF)
observed at $E_{cyc}\sim45$\,keV in the X-ray spectrum of the pulsar. Unlike
many other sources, \src exhibits no significant variation of the line energy
with luminosity, suggesting that the line is formed close to the surface of the
neutron star \citep{Caballero13}.

The variability of the source is not limited to coherent pulsations and
outbursts. During the 1994 giant outburst, quasiperiodical oscillations (QPOs)
with frequency in the range 27--72~mHz have been observed in the source's hard
(20-50\,keV) X-ray light curve \citep{Finger96}. The QPO frequency was found to
be correlated with flux, as was recently confirmed by \emph{Fermi}~GBM
observations in similar energy range \citep{ca12}. This allowed the association
of the QPOs with the inner edge of the accretion disk and the magnetosphere
boundary \citep{Finger96, ca12}, although there are some problems with this
interpretation.

Another manifestation of aperiodic variability in \src, is the broken power-law
shape of the noise power spectrum, a feature common to many accreting pulsars
\citep{fbreak0,revnivtsev,Tsygankov2012}. \cite{revnivtsev}, using RXTE
observations carried out during the source outburst of 2009, were able to show
that similarly to QPOs, the frequency of the break is correlated with flux, and
proportional to the Keplerian frequency at the inner edge of the accretion disk
where the disk is disrupted by the magnetosphere. The power spectrum was
interpreted in the framework of the perturbation-propagation model proposed by
\cite{lyubarskii_flicker} and \cite{churazov}, in which the fluctuations in
accretion rate are associated with fluctuations in the accretion disk alpha
parameter $\alpha$, occurring on Keplerian timescales. This model predicts a
$P\sim f^{-1}$ power spectrum for the resulting accretion rate fluctuations,
consistent with power spectra of accreting pulsars below the break
\citep{revnivtsev}. Above the break, instead, the observed spectra are steeper
with $P\sim f^{-2}$, which corresponds to the integrated white noise power
spectrum with uncorrelated rate increments and can be naturally associated with
free-falling plasma within the magnetosphere. The break in the power spectrum
can therefore be associated with the disruption of the accretion disk, and the
fact that this happens at a distance proportional to the magnetospheric radius
strongly supports this interpretation.

Because it is a transient source, \src is best studied at high luminosities
during the outbursts associated with passage of the neutron star through the
circumstellar decretion disk of the \emph{Be} primary. When the neutron star
leaves the vicinity of periastron and/or the decretion disk recedes, the
density of the plasma surrounding the neutron star drops, and the source enters
the so-called quiescence period with drastically reduced (if not ceased) X-ray
flux \citep{Reig11, Rothschild13}.

Accretion onto rotating magnetized neutron stars is expected to be centrifugally
inhibited below a certain accretion rate \citep{Illarionov:1975p2044}, and it
was long assumed that this is indeed the case for Be systems in quiescence.
However, with the increased sensitivity of X-ray instruments, pulsed X-ray flux
was detected from several sources, including 1A 0535+262, outside of outbursts,
during states at luminosities below the critical known as the quiescent
state \citep{Motch91, campana02, orlandini04, rutledge07, Rothschild13}. In
\src the quiescent flux is almost certainly powered by accretion \citep{negu00,
ikhsanov01a}, although it is not clear how the matter leaks through the
magnetosphere. To clarify the origin of its quiescent emission and the
accretion mechanisms behind it, we observed \src with XMM-Newton for 50\,ks to
investigate in detail the spectral and timing properties of the source in
quiescence. The results are presented and discussed in this paper.

\section{Observations and data analysis} We observed \src with
\emph{XMM-Newton} for 50\,ks on Feb. 28, 2012. At the time of observation, the
neutron star was close to apastron, and the source had shown no outbursts in
the two preceding orbital cycles. Therefore, \src was certainly deep in
quiescence. We also used the archival \emph{Suzaku} observation (ID.~100021010)
performed at the end of a normal outburst in Sep~2005 as a reference to
compare the quiescent and outburst spectra in the 0.2-12 keV energy range. In
addition, we used all available observations of \src by the Proportional
Counter Array (PCA) onboard the Rossi X-ray Timing Explorer (RXTE) to
investigate the aperiodic variability of the source at higher fluxes. The data
reduction was carried out using the
\emph{XMM}~SASv12.0\footnote{http://xmm.esa.int/sas} and
HEASOFT~6.12\footnote{http://heasarc.gsfc.nasa.gov/docs/software/lheasoft}
software packages.

\begin{figure}[t]
    \centering
        \includegraphics{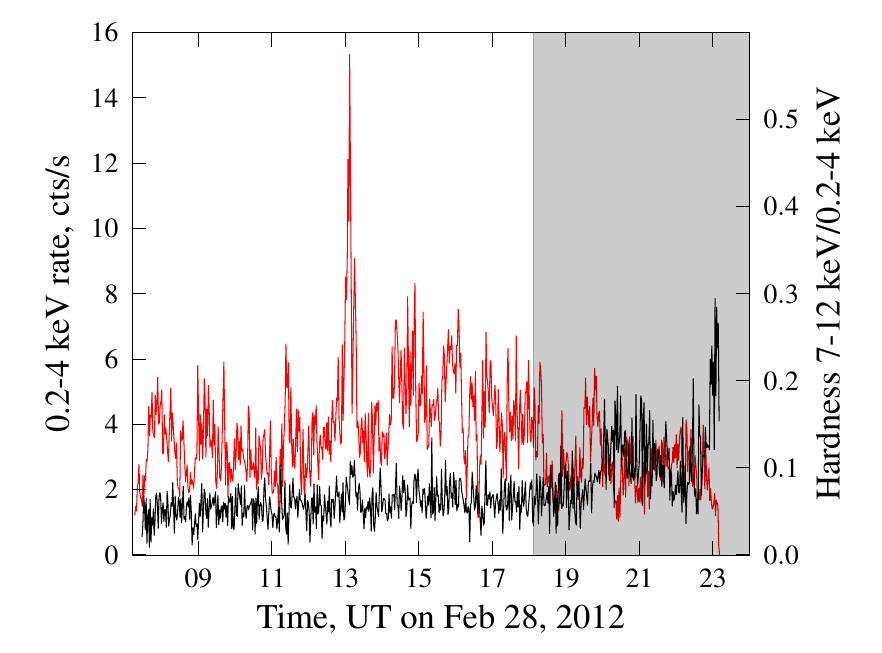}
    \caption{Observation-long background-subtracted EPIC-PN light curve
    in the soft 0.2-4 keV energy range with time bin of 100\,s (red), and
    hardness ratio for rates in 7-12\,keV and 0.2-4\,keV bands (black).
    Shaded area shows the part of the observation affected by enhanced
    particle background discarded in spectral analysis.}
    \label{fig:obslc}
\end{figure}

In \emph{XMM} data, the source was clearly detected with all three instruments
with an average combined count-rate of about 5~counts/s and a factor of four
more in a flare-like episode halfway through the observation, which corresponds
to source flux of $\sim2.7\times10^{-11}{\rm\,erg\,cm}^{-2}{\rm\,s}^{-1}$ and
luminosity of $\sim1.3\times10^{34}{\rm\,erg\,s}^{-1}$ assuming a distance of
2\,kpc. The observation-long EPIC-PN light curve of \src is presented in
Fig.~\ref{fig:obslc}. The end of the observation ($\sim18$\,ks) is severely
affected by particle background, so we discard it in spectral analysis and only
use it for timing.

\subsection{Timing analysis} The search for pulsations and a detailed study of
pulse-profiles was one of the main goals of our observation. For timing
analysis, the photon arrival times were corrected for the orbital motion in the
solar system and in the binary system using the ephemeris by \cite{Finger96}
with the adjusted epoch provided by the Fermi GBM Pulsar
project\footnote{http://gammaray.nsstc.nasa.gov/gbm/science/pulsars}. We then
extracted the light curve in the 0.2-12\,keV energy band with 1\,s time bins
using the data combined from all three EPIC cameras. For this light curve, the
Lomb-Scargle periodogram reveals a single highly significant peak at
$\sim103.28$\,s coincident with the spin period of a neutron star, so we
confirm that the X-ray flux is pulsed in quiescence.

To refine the pulse period and search for possible period evolution during the
observation, we also performed a pulse-phase coherent timing analysis
\citep{Nagase:1982p2720, Doroshenko:2010p3661}. We find that the data are
consistent with a constant period of $P=103.286(6)$\,s (all uncertainties are
reported at a $1\sigma$ confidence level unless stated otherwise), and no pulse
period derivative is required.

\begin{figure}[t]
	\centering
		\includegraphics{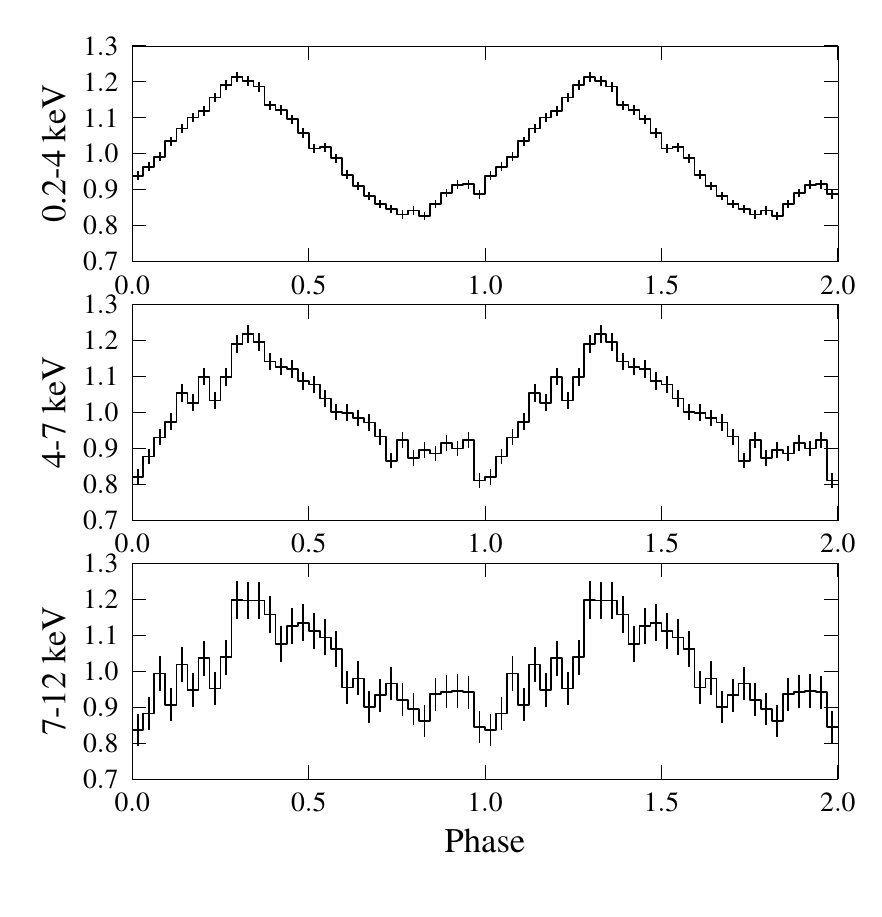}
        \caption{Normalized energy-resolved pulse profiles folded with
        best-fit period using combined PN and MOS background subtracted light curves (background rate <0.7\% in all cases).}
	\label{fig:pp}
\end{figure}

Using the refined period value, we folded the light curves in several energy
ranges to obtain the pulse-profiles presented in Fig.~\ref{fig:pp}. The pulse
shape is sine-like and almost constant with energy, in line with earlier reports
for the observations of the source in quiescence \citep{negu00,Mukherjee05,
Rothschild13}. As noted by \cite{Rothschild13}, in comparison with outburst
observations, this simple shape resembles hard rather than soft pulse profiles.

Although we were unable to detect any change in pulse period during the
\emph{XMM-Newton} observation, the source is known to spin-down between the
outbursts. To investigate the spin evolution of the source as a function of
flux both in outburst and in quiescence, we used the \emph{Fermi}~GMB pulse
frequency history provided for \src by the GBM~Pulsar archive$^3$, which
contains the measurements of the frequency and pulsed signal amplitude over a
set of one- to two-day day intervals, when the source was in outburst.

To investigate changes of the source spin during outbursts, we compared the
GBM~Pulsar frequency values reported for adjacent time intervals averaging the
left and right derivatives for each point \citep{Doroshenko11}. To estimate the
uncertainty for the frequency derivative, we used the bootstrap technique,
repeating this procedure 1000 times for frequencies normally distributed for
each point within the reported GBM uncertainties, and averaged the results.

To determine the average spin-down rate in quiescence, we compared the
frequencies (in the GBM~Pulsar archive) at the end and at the onset of each
pair of consequent outbursts observed by the GBM (see also \citealt{ca12} for
more details). We also compared the frequency measured by GBM at the end of the
most recent outburst with the one derived from the \emph{XMM} data. By
averaging these values, we estimated the spin-down between the outbursts to be
$\dot{\nu}=-6.8(1.4)\times10^{-14}{\rm\,Hz\,s}^{-1}$. The relatively large
uncertainty reflects the scatter of individual measurements.

To study the luminosity dependence of the source spin, one also needs a
reliable estimate of the source flux, which is not trivial in the case of the
\emph{Fermi}~GBM, which can only measure the amplitude of pulsed flux. To
determine the energy flux and the accretion rate from the GBM amplitudes, we
used the available RXTE observations contemporary with the GBM measurements in
order to calibrate observed pulsed amplitudes with the source flux derived from
spectral fits. The flux and GBM amplitude were found to be well correlated with
a rank correlation coefficient of $r\sim0.997$ and $f_{3-20}[{\rm
erg\,cm}^{-2}{\rm\,s}^{-1}]\sim10^{-8}A$, where $A$ is the pulsed flux
amplitude measured by GBM. We use this dependence later on to estimate the flux
for \emph{Fermi}~GBM observations.

\begin{figure}[t]
    \centering
        \includegraphics{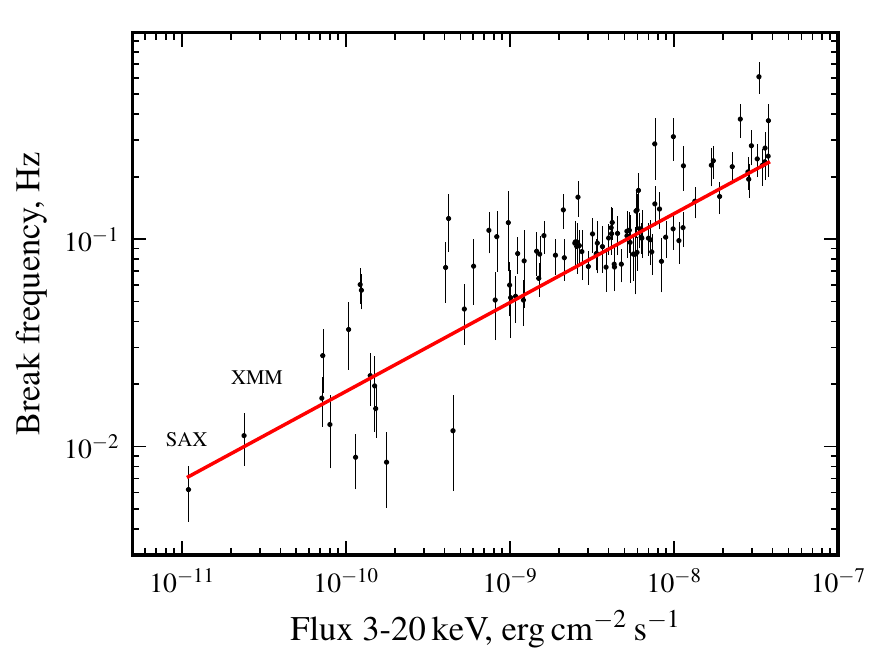}
        \includegraphics{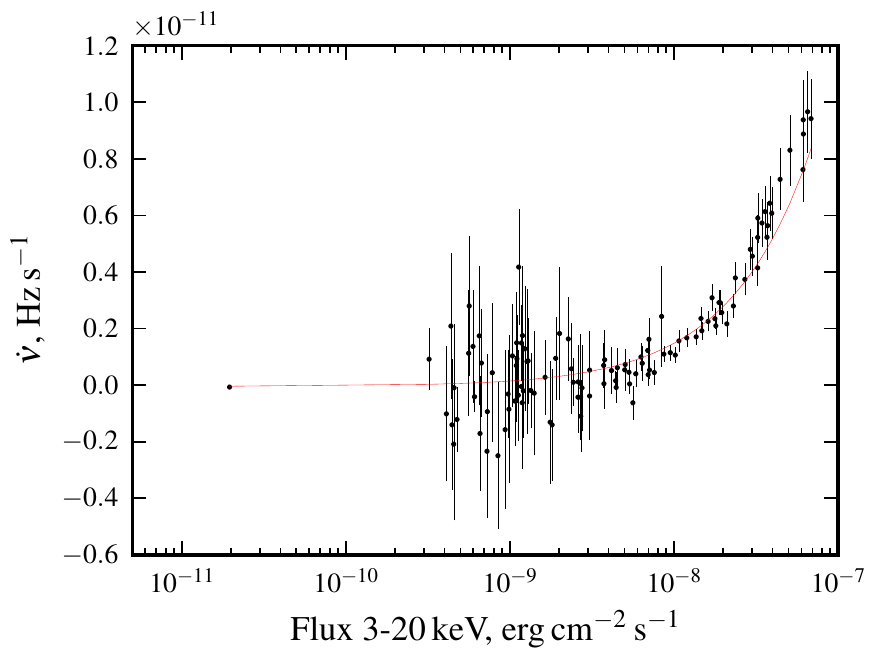}
        \caption{Break frequency (top panel) and spin frequency
        derivative (bottom) as a function of flux and the best-fit model
        prediction for the joint fit as described in the text (red lines).}
    \label{fig:fbreak_fdot}
\end{figure}

To estimate the average quiescent flux, we used the best-fit spectral model for
XMM data to scale the quiescent fluxes reported in the literature
\citep{Motch91,Mukherjee05,negu00} to the same 3-20\,keV energy range.
Averaging the resulting values (also taking into account the value obtained for
our XMM observation) implies a quiescent 3-20\,keV flux of
$f_{3-20}\sim2\times10^{-11}{\rm\,erg\,cm}^{-2}{\rm\,s}^{-1}$. The dependence
of the pulse frequency derivative on flux is presented in
Fig.~\ref{fig:fbreak_fdot} (the lowest flux point corresponds to our estimate
of average flux and frequency derivative, and the others correspond to the
\emph{Fermi} data).

Motivated by the analysis of aperiodic variability in \src presented by
\cite{revnivtsev}, we extended this study to our observation. For this
analysis, we again used the full energy range (0.2-12\,keV) light curve with a
time bin of 1\,s, combining the data from both MOS and PN cameras. The break
around the pulse frequency is immediately apparent in the resulting power
spectrum, although, as discussed by \cite{revnivtsev}, the contribution of the
pulsed flux has to be removed to constrain the frequency of the break.

To this end, \cite{revnivtsev} subtracted the average pulse profile from each
gap-free part of the light curve (i.e., each ten pulses, which corresponds to
the average usable observation duration per RXTE orbit) before constructing the
power spectra. This approach, however, may potentially act as a low-pass filter
for frequencies below the inverse size of the window used to subtract the
pulsed component. Moreover, because of the intrinsic variability of the pulse
profiles, this window should not be too large otherwise the pulsed component
cannot be subtracted completely. While it is not a concern for the RXTE light
curves which are interrupted each orbit, our light curve contains no gaps and a
different approach must be used to extend the power spectrum to lower
frequencies. The simplest option is to ignore the frequencies corresponding to
the harmonics of the pulse frequency of the pulsar when fitting the power
spectrum, which is the conservative approach we follow. To fit the resulting
power spectrum, we used a simple broken power-law model in the form $P(f) =
N_{signal}(f/f_{break})^{\alpha_i}+N_{noise}$. Here, $f_{break}$ is the break
frequency, $\alpha_1>-2$ and $\alpha_2=-2$ are power indices below and above
the break, and $N_{signal, noise}$ are power-law and noise amplitudes. The
choice of this model over of the King profile used by \cite{revnivtsev} is
motivated by the fact that the King profile, in addition to the two power laws
in the spectrum, also includes a smooth transition region, which is not evident
in the observed spectra and may affect the estimated break values.

These methodological differences complicate a direct comparison between the
break frequency values reported by \cite{revnivtsev} and the ones resulting
from our \emph{XMM} measurement. We therefore re-analyzed all available
\emph{RXTE} observations (including the data used by \citealt{revnivtsev})
following the same procedure as for the \emph{XMM} data. We only considered the
PCA data for the analysis. We estimated the power spectrum for each RXTE
observation by averaging the power spectra of continuous data segments within
it. For further analysis, we only considered observations where the frequency
break could be significantly detected.

The X-ray flux is derived for each observation from the observation-long X-ray
spectrum fitted with an absorbed cutoff power-law model. In addition, we
analyzed a single \emph{BeppoSAX} observation of the source in quiescence, when
pulsations were still detected \citep[observation C in][]{Mukherjee05}. This
observation contains, however, many data gaps due to the occultation of the
source by the Earth, making it difficult to constrain the power spectrum at low
frequencies and, consequently, the break frequency. The results are presented
in Fig.~\ref{fig:fbreak_fdot}. The correlation of the break frequency with flux
reported by \cite{revnivtsev} is confirmed and, moreover, seems to extend to
the lowest fluxes where the pulsed flux from the source is still detected, with
both \emph{BeppoSAX} and \emph{XMM-Newton} measurements following this trend.
We note that in quiescence the break frequency approaches the spin frequency of
the neutron star below which the accretion shall be inhibited centrifugally,
which may explain the non-detection of pulsations at lower fluxes in
\emph{BeppoSAX} observations \citep[observations A and B in][]{Mukherjee05}.

\subsection{Spectral analysis} Broadband coverage, unfortunately lacking in our
observation, is essential for a robust modeling and proper interpretation of
the components of the continuum X-ray spectra of accreting pulsars. For
instance, \cite{Naik08} report a blackbody-like component with a temperature of
$\sim1.4$\,keV, which is only required when the broadband spectrum is
considered, while the fit of the soft spectrum (0.2-12\,keV) alone does not
require two components. The comparison of quiescent and outburst spectra in
similar energy ranges, however, could clarify whether the spectrum changes or
not. As a reference for this comparison, we use a $\sim43$\,ks long
\emph{Suzaku} observation (ID.~100021010) of \src carried out at the end of a
normal outburst in Sep.~2005 and analyzed in detail by \cite{Naik08}. We
restrict the analysis to X-ray Imaging Spectrometer (XIS) data, which has a
similar energy range to the \emph{XMM} EPIC cameras and combine the spectra
from the three front illuminated XIS units.

The effective exposure after rejection of periods with high particle background
is about 33 and 37\,ks for \emph{XMM} PN and MOS, respectively. We extracted
spectra and background separately for each camera, to account for differences
in energy response, and we then fitted them simultaneously. In all cases
spectra were re-binned to contain at least one hundred photons per bin. In
addition, we separately considered the large flare which can be seen in
Fig.~\ref{fig:obslc} and lasts $\sim1.3$\,ks to probe for flux dependent
spectral change within the \emph{XMM} observation. We note that similar events
have been reported by \cite{hill07} based on the \emph{INTEGRAL} observations
of the source. Flaring is probably normal for \src in quiescence.

\begin{figure}[t]
    \centering
        \includegraphics{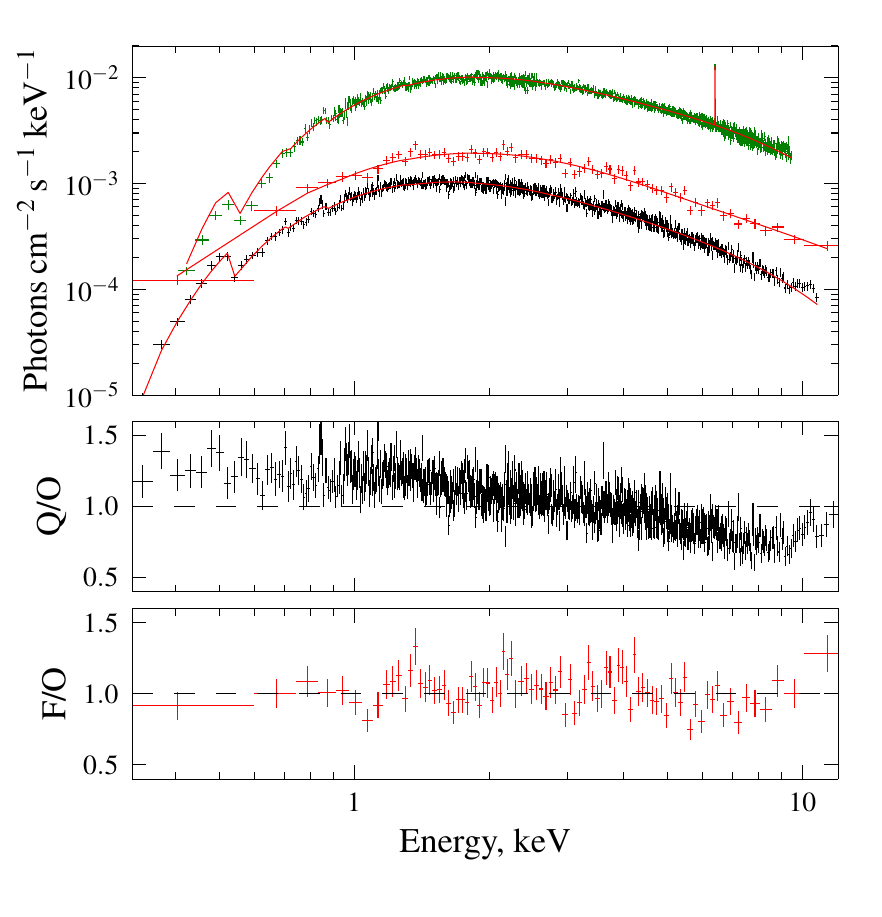}
    \caption{
    Best-fit outburst, flare, and quiescent spectra (top to bottom in top
    panel) using the Suzaku XIS and the XMM PN data, and the ratio of
    \emph{XMM} average (Q) and flare (F) spectra to best-fit model for Suzaku
    XIS outburst spectrum (O) with adjusted absorption column and normalization.
    }
    \label{fig:spe}
\end{figure}

Both \emph{XMM} and \emph{Suzaku}~XIS spectra are well fitted, either with
single component models commonly used for the source, namely the absorbed
cutoff power law, and the Comptonization model \citep[$CompTT$ in \emph{Xspec},
see][]{Titarchuk:1994p2324}. In the \emph{XMM} data, there is no evidence of
the fluorescence iron line marginally detected in the \emph{Suzaku} spectrum
\citep[see also][]{Caballero13}. We wish to note that both models are flexible
enough to explain the blackbody-like soft component reported by
\cite{Mukherjee05} and \cite{Naik08} in the \emph{XMM} energy range. The
unfolded spectrum and best-fit model parameters for \emph{XMM} average, flare,
and the \emph{Suzaku} XIS spectra are presented in Fig.~\ref{fig:spe} and
Table~\ref{tab:spe}. In XMM observation the source flux in the 0.2-12\,keV
energy range is $2-6\times10^{-11}{\rm erg\,cm}^{-2}{\rm\,s}^{-1}$ (on average
and during the flare, respectively, i.e., at least a factor of six less than in
the \emph{Suzaku} observation), which corresponds to an average luminosity of
$\sim1.3\times10^{34}{\rm erg\,s}^{-1}$ assuming a distance of 2\,kpc. Except
for a difference in the inferred absorption column, the spectra are similar in
shape, although the source gets softer as the flux decreases. To highlight
these changes, in Fig.~\ref{fig:spe} ratios of the \emph{XMM} average and flare
spectra to the best-fit model for the outburst spectrum with adjusted
normalization and absorption column fixed to the value measured by \emph{XMM}
are plotted. The outburst and flare spectra seem to have a similar shape,
whereas the average quiescent \emph{XMM} spectrum is slightly softer. Both
models essentially describe the same Comptonization spectrum emerging from the
hotspots on the surface of neutron star, so we refer to the \emph{CompTT} for
the rest of the paper because of the more straightforward interpretation of the
parameters in this model.

\begin{figure}[t]
    \centering
        \includegraphics{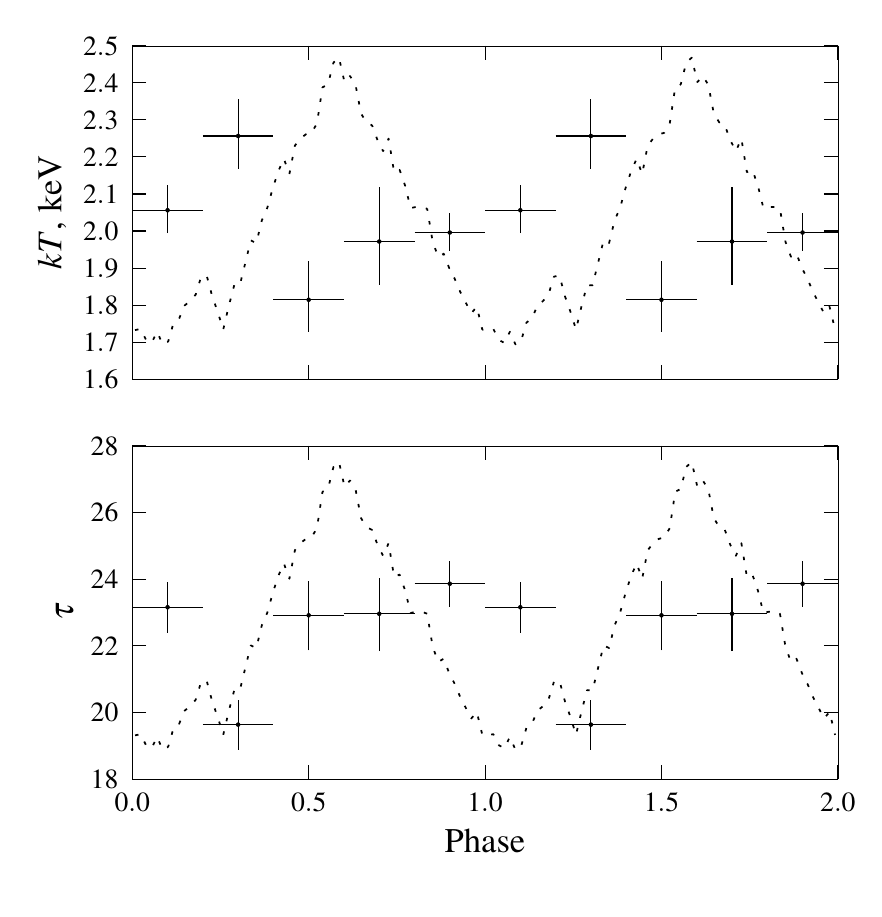}
        \caption{Variation of the continuum parameters with pulse phase.
        The scaled pulse-profile in the 0.2-12\,keV energy range is also
        plotted for reference.}
    \label{fig:phres}
\end{figure}

The energy-related changes of the pulse profile shape suggest a pulse-phase
dependence of the spectrum. To investigatethes we analyzed separately the
\emph{XMM} spectra in five equally spaced phase bins separately. The zero phase
and binning were chosen in order to match the main features of the pulse
profile, namely the hard shoulder, the dip after it, and the rising and falling
shoulders of the main peak. The best-fit results for the pulse-phase resolved
spectra with the \emph{CompTT} model are presented in Fig.~\ref{fig:phres} and
Table~\ref{tab:spe}. There seems to be no strong phase dependence of spectral
parameters, with the exception of the dip, where the spectrum significantly
softens owing to a drop in the optical depth. This behavior is opposite to what
one could expect due to the obscuration of the emission region by the accretion
flow, and instead we seem to observe more directly the softer seed photons. One
possibility is that we are looking at the polar cap directly through a hollow
accretion column. This agrees qualitatively with the results of pulse profile
decomposition carried out for the source by \cite{Caballero2011}.

\begin{table*}
    \begin{center}
    \begin{tabular}{lcccccc}
\multicolumn{7}{c}{Pulse-phase average}\\

\hline
\hline
\noalign{\vskip 1mm} 
                                 & \multicolumn{3}{c}{\emph{cutoffpl}} & \multicolumn{3}{c}{\emph{CompTT}} \\
                                 & \emph{XMM}, all & \emph{XMM}, flare & \emph{Suzaku} & \emph{XMM}, all & \emph{XMM}, flare & \emph{Suzaku}\\
      \cline{2-7}
      \noalign{\vskip 2mm} 
 $N_{\rm H}$\, atoms\,cm$^{-2}$  & 0.380(7)   & 0.38(3)   & 0.516(8)  &  0.218(8)  & 0.20(3) &0.275(9)\\
  $\Gamma$/$\tau$                & 0.38(2)    & 0.2(1)    & 0.38(2)   &  23.3(5)  & 22(2) & 20.7(4) \\
  $E_{fold}$/($T_0, kT$),\,keV          & 4.2(1)     & 4.1(5)    & 5.6(1)    &  0.44(1), 1.98(3)  & 0.52(4), 2.3(3) & 0.541(9), 2.57(5)\\       
  $\log_{10}{F}$                 & -10.530(2) & -10.19(1) & -9.410(2) &  -10.576(2)  & -10.221(1) & -9.442(2)\\
  \noalign{\vskip 1mm}
  \cline{2-7}
  \noalign{\vskip 1mm}
   $\chi^2_{red}/DoF$                & 1.15/1387       &  1.4/1387      & 1.17/1387      &  1.11/1387  & 1.4/1387 & 1.18/1387\\
\multicolumn{7}{c}{Pulse-phase resolved (with \emph{CompTT})}\\
\hline
\hline
\noalign{\vskip 1mm} 
                            Phase interval &         &            0-0.2 &    0.2-0.4 &   0.4-0.6 &   0.6-0.8 &    0.8-1.0\\
\hline
\noalign{\vskip 2mm} 
                                $kT$,\,keV &     &          2.06(7) &     2.3(1) &    1.8(1) &    2.0(1) &    2.00(5)\\
                                    $\tau$ &     &          23.2(8) &    19.6(8) &     23(1) &     23(1) &    23.9(7)\\
        $\log_{10}{F}$ & &       -10.615(5) & -10.622(5) & -10.57(1) & -10.54(2) & -10.589(4)\\
        \noalign{\vskip 1mm}
        \cline{3-7}
        \noalign{\vskip 1mm}
        $\chi^2_{red}/DoF$&  & \multicolumn{5}{c}{1.11/1143} \\
    \end{tabular}
    \end{center}
    \caption{Best-fit spectral parameters for average and flare
    \emph{XMM} and outburst \emph{Suzaku} spectra fitted with
    \emph{cutoffpl} and \emph{CompTT} models, and for pulse-phase
    resolved \emph{XMM} spectra fitted with the \emph{CompTT} model. 
    The flux logarithm $\log_{10}{F}$\,[erg\,cm$^{-2}$\,s$^{-1}$] is
    the unabsorbed source flux in the 0.2-12\,keV energy range.
    Phase-averaged spectra were fit simultaneously with common absorption
    column and seed photon temperature, which converged to phase-averaged values.
    The provided $\chi^2_{red}$ is for this combined fit.
    }
    
    \label{tab:spe}
\end{table*}

\section{Discussion} The analysis of the \emph{XMM-Newton} observation of \src
in quiescence and the comparison of the results with data from other satellites
has revealed a phenomenology similar to that of the source in outburst:

\begin{itemize}
    \item The source is highly variable, with the flux increasing by a factor
    of ten during the flares. The average flux in quiescence also appears to
    vary when compared to the quiescence fluxes reported with different
    instruments.
    
    \item The X-ray spectrum, particularly during the higher flux flare
    episode, is similar to that in outburst (within the \emph{XMM} energy
    range).

    \item Although the pulse profile is simplified in quiescence, some features
    typical of outburst profiles can still be seen. The amplitude of the pulsed
    flux is also similar.
    
    \item The noise power spectrum exhibits the same broken power-law shape as
    at higher fluxes, and the break frequency follows the same correlation with
    flux as at higher luminosities.
    
    \item The neutron star spins down in quiescence at an average rate of
    $\dot{\nu}=-7\times10^{-14}{\rm\,Hz\,s}^{-1}$.
\end{itemize}

First of all, these results, in line with previous reports \citep{Motch91,
ikhsanov01a}, confirm that the observed emission is powered by the accretion of
plasma from the primary onto the neutron star. \cite{ikhsanov01a} discussed why
other mechanisms fail to explain the observed phenomenology, and detection of
the break in the quiescent power spectrum may be the first direct evidence that
the observed emission is indeed powered by accretion onto the magnetized
neutron star. The question is, however, how the accretion proceeds.

In the case of the spherically symmetric accretion geometry, observed quiescent
fluxes imply that the stationary accretion should be centrifugally inhibited
\citep{Illarionov:1975p2044, Motch91}. The magnetospheric radius
$R_m={(\mu^2/(2\dot{M}\sqrt{2GM}))}^{2/7}\sim 8\times10^{9}\,{\rm cm,}$ exceeds
the co-rotation radius $R_c=(GM/\omega^2)^{1/3}\sim 4\times10^9$\,cm by a
factor of two, so the magnetosphere rotates at a super-Keplerian velocity and,
therefore, accretion is not possible. In the equation, $\mu$ is the dipole
magnetic moment of the neutron star, $M$ is the mass, and $\dot{M}$ is the
accretion rate. In principle, plasma may still leak through the magnetosphere
owing to various instabilities as discussed by \cite{elsner77,Perna:2006p2027}
and \cite{Bozzo:2008p2039}. For instance, for quiescent emission in \src,
\cite{ikhsanov01a} considered instabilities associated with field line
reconnection, concluding that the observed quiescent X-ray fluxes are
compatible with the expected plasma leak rate, so quasispherical accretion is
certainly not excluded.

On the other hand, \cite{ikhsanov01a} notes that for disk accretion the
critical luminosity is lower, and, therefore, no additional instabilities are
required to enable stationary accretion in this case. \cite{chichkov_disk}
considered this possibility and concluded that, because of the fast rotation of
the primary in \src, the accreting matter might have sufficient angular
momentum to form an accretion disk around the neutron star at all orbital
phases. Direct evidence of an accretion disk around the neutron star was
reported by \cite{Giovannelli07} based on the velocities derived from the
observed doubling in the He~I emission lines in the optical spectrum of the
source. We note that at the time of the observation in Oct~1999, the source
remained X-ray quiet, although the neutron star was close to the periastron.

Moreover, the onset of the normal outburst in Sep~2005 was accompanied by
flaring activity \citep{caballero_thesis}, which \cite{Postnov:2008p348}
associated with magnetospheric instabilities emptying the inner regions of the
accretion disk. This implies that an accretion disk was already in place at the
beginning of outburst. \cite{Postnov:2008p348} also noted that a similar
mechanism might be responsible for short flare events like the one reported by
\cite{hill07} (which appear to be very similar to the flare observed in our
\emph{XMM} observation).

It is interesting to note that the spin evolution of \src is also consistent
with disk accretion. \cite{caballero_thesis} has shown that the spin evolution
of \src in outburst seems to be in agreement with the predictions of the torque
model by \cite{Ghosh:1979p1676} for the case of disk accretion. To verify if
this is the case, and whether the same model can describe the spin-down of \src
between the outbursts, we have applied the same model to the observed
spin-frequency derivative - flux dependence obtained as described above.

We considered the distance to the source, necessary to convert flux to
accretion rate as the only free parameter. The uncertainty in accretion
efficiency was assumed to be around ten percent and may also be attributed to
uncertainty in distance. Other parameters of the model were fixed. The mass and
radius of the neutron star were assumed to be $M=1.5M_\odot$ and
$R_{ns}=13$\,km, respectively \citep{Suleimanov11}, and the magnetic field was
fixed to the value derived from the energy of the cyclotron line observed in
\src. As discussed by \cite{Caballero13}, the cyclotron line energy is observed
at $E\sim46$\,keV and does not change significantly with the luminosity. This
implies that it most likely originates in the vicinity of the neutron star
surface and, therefore, does provide an estimate of the magnetic field at the
surface. In this case, $B/10^{12}{\rm[G]}=(1+z)/11.57\times46{\rm
[keV]}\sim4.9\times10^{12}$\,G, where $z$ is the gravitational redshift.

Even with these restrictions, the spin evolution of \src is well described by
the model (see Fig.~\ref{fig:fbreak_fdot}). The best-fit distance was found to
be $d=1.85(2)$\,kpc, in excellent agreement with existing measurements. For
this value of the distance, the model predicts
$\dot{\nu}\sim-3.8\times10^{-14}{\rm\,Hz\,s}^{-1}$ at average quiescence
luminosity, which is in reasonable agreement with observations if one takes
into account the uncertainty in the average flux value. We conclude, therefore,
that the \cite{Ghosh:1979p1676} model for disk accretion describes the spin
frequency evolution well, both during and between the outbursts.

The presence of a break in the noise power spectrum, and the flux - break
frequency'' correlation revealed by \emph{RXTE} observations in outburst,
\cite{revnivtsev}, can be naturally interpreted as the disruption of the
accretion disk at and by the magnetosphere. In this work, we have shown that
the noise power spectrum at quiescent fluxes, obtained from our \emph{XMM}
observation has a similar shape as well. Moreover, we have extended the
observed break frequency vs. flux correlation observed at higher fluxes during
the outbursts to lower fluxes, which also suggests that the accretion geometry
does not change in quiescence and that the accretion still proceeds from the
disk.

\cite{revnivtsev} have associated the break frequency with the Keplerian
frequency at a radius proportional to the magnetosphere radius $f_b=\nu_{\rm
K}(k\cdot R_m)=(GM)^{1/2}(k\cdot R_m)^{-3/2}/2\pi$. The magnetospheric radius
as a function of the flux can be constrained from the spin evolution of the
neutron star, so we can find $k$ (and the source distance) simultaneously,
fitting the spin frequency derivative and the break frequency as functions of
flux. The best-fit values are $d=1.85(3)$\,kpc, and $k=0.52(1)$. We note that
the value of $k$ matches the predictions of the \cite{Ghosh:1979p1676} model,
so spin evolution and the break frequency changes are described
self-consistently.

Finally, an independent estimate of the plasma velocity in the disk may be
obtained from high-resolution X-ray spectroscopy. \cite{Reynolds10}, using
\emph{Chandra} high-resolution grating spectra of two observations performed at
the end of the 2009 outburst, observed the broadening of $Fe~{\rm K}_{\alpha}$
lines in the spectrum of \src, measuring velocities of $\sim5090(1000)
{\rm\,km\,s}^{-1}$ and $\sim4100(1000) {\rm\,km\,s}^{-1}$, respectively. The
two \emph{Chandra} observations were almost simultaneous with \emph{RXTE}
observations 94323-05-03-03 and 94323-05-03-06. For these observations, we
derive a flux in the 3-20\,keV energy range of
$1.63\times10^{-8}{\rm\,erg\,cm}^{-2}{\rm\,s}^{-1}$ and $9.7\times10^{-9}{\rm
erg\,cm}^{-2}{\rm\,s}^{-1}$ corresponding to break frequencies of 0.16 and
0.13\,Hz, respectively, and to azimutal velocities in the disk of
$\upsilon_\phi=(2\pi \nu_kGM)^{1/3}\sim5900{\rm\,km\,s}^{-1}{\rm
and~5500\,km\,s}^{-1}$. These results are, within the uncertainty, consistent
with the observations even without taking into account the unknown inclination
of the disk.

Several independent lines of evidence suggest, therefore, that in \src, the
accretion disk around the neutron star powers the accretion, not only in
outbursts, but also in quiescence. A similar scenario may be realized in other
\emph{Be} systems as well, and may be very important for understanding the spin
evolution of enclosed neutron stars and the physics of quasiperiodic outbursts.
We would like to mention here the model proposed by \cite{Syunaev77} to explain
irregular outbursts observed from transient neutron stars, where the authors
argued that interaction of the accreting plasma with the magnetosphere may
partially inhibit the accretion and lead to formation of the so-called ``dead''
disks, where the disk accumulates matter until it becomes unstable, resulting
in an outburst. While it is clear that in \src outbursts are closely related to
the extension of the circumstellar disk of the primary, stability of the
accretion disk around the neutron star may also be important. In particular, it
may serve as a trigger for the onset of outbursts setting their exact timing,
which would explain a relatively loose connection of the observed outbursts to
periastron.

\section{Conclusions} To clarify the origin of the quiescent X-ray emission in
\src, we proposed, and were granted, a 50\,ks observation with
\emph{XMM-Newton}, which was carried out on Feb.~28, 2012, when the neutron
star was close to apastron. The system had exhibited no outburst activity since
two orbital cycles preceding the observation, and thus was in X-ray quiescence.
Similarly for earlier reports \citep{Rothschild13}, we were able nevertheless
to detect pulsed flux from the source at a level of
$\sim2.6\times10^{-11}\,{\rm erg\,cm}^{-2}{\rm\,s}^{-1}$ and to perform
detailed spectral and timing analysis. The derived pulse period of
103.286(6)\,s implies a spin-down with respect to the most recent outburst.
Data quality unveiled for the first time a simplified, but still complex, shape
of the pulse profiles in quiescence with certain features resembling the shapes
of the outburst profiles.

We found that within the \emph{XMM-Newton} energy range the X-ray spectrum is
similar to that observed in outbursts, and even more so for the flare-like
event with a flux factor of $\sim3$ higher than average, which occurred during
the observation. Taking into account similarities of the spectral and timing
properties observed in quiescence and outbursts, it is clear that the observed
flux is powered by accretion. We have also investigated the pulse-phase
dependence of the spectrum, and found that the continuum parameters remain
relatively stable with the pulse phase except for a sharp dip, which can also
be traced in the outburst pulse profiles. There the spectrum softens, which can
be interpreted as a decrease in optical depth of the Comptonizing region as,
for instance, when the observer looks through a hollow column.

We also investigated the aperiodic variability in the source, and found that in
quiescence the noise power spectrum shows the same broken power-law shape as
during outburst observations, and moreover, the break frequency follows the
same correlation with flux. We argue that this could be explained if the
accretion geometry is similar in both cases, i.e., if the observed X-ray
emission is powered by an accretion disk around the neutron star. Although
quasispherical accretion cannot be ruled out, we find that this scenario allows
us to consistently explain the observed break and spin frequency dependence on
flux, and to discuss other previously published observational arguments which
also support it.

It was proposed by \cite{Syunaev77} that centrifugal inhibition might lead to
the formation of the so-called dead disks, where the matter is accumulated
until the disk or magnetosphere becomes unstable. If this scenario is realized
and the accretion disk is indeed ever present in \src, this may have important
consequences for understanding the mechanisms that trigger the outbursts in
transient neutron star \emph{Be} systems, which may be related both to the size
of circumstellar disk of the primary, and to the stability of the accretion
disk and magnetosphere as discussed by \cite{Postnov:2008p348}.

\begin{acknowledgements} The authors thank M.~Revnivtsev for many useful
discussions and help with the power spectrum analysis. VD and AS thank the
Deutsches Zentrums for Luft- und Raumfahrt (DLR) and Deutsche
Forschungsgemeinschaft (DFG) for financial support (grant DLR~50~OR~0702).
\end{acknowledgements}

\vspace{-0.3cm}\bibliography{auto_clean}	\vspace{-0.3cm}
\begin{thebibliography}{39}
\expandafter\ifx\csname natexlab\endcsname\relax\def\natexlab#1{#1}\fi

\bibitem[{{Bozzo} {et~al.}(2008){Bozzo}, {Falanga}, \&
  {Stella}}]{Bozzo:2008p2039}
{Bozzo}, E., {Falanga}, M., \& {Stella}, L. 2008, \apj, 683, 1031

\bibitem[{{Caballero}(2009)}]{caballero_thesis}
{Caballero}, I. 2009, PhD thesis, IAAT University of Tuebingen

\bibitem[{{Caballero} {et~al.}(2011){Caballero}, {Kraus}, {Santangelo},
  {Sasaki}, \& {Kretschmar}}]{Caballero2011}
{Caballero}, I., {Kraus}, U., {Santangelo}, A., {Sasaki}, M., \& {Kretschmar},
  P. 2011, \aap, 526, A131

\bibitem[{{Caballero} {et~al.}(2013){Caballero}, {Pottschmidt}, {Marcu},
  {Barragan}, {Ferrigno}, {Klochkov}, {Zurita Heras}, {Suchy}, {Wilms},
  {Kretschmar}, {Santangelo}, {Kreykenbohm}, {F{\"u}rst}, {Rothschild},
  {Staubert}, {Finger}, {Camero-Arranz}, {Makishima}, {Enoto}, {Iwakiri}, \&
  {Terada}}]{Caballero13}
{Caballero}, I., {Pottschmidt}, K., {Marcu}, D.~M., {et~al.} 2013, \apjl, 764,
  L23

\bibitem[{{Camero-Arranz} {et~al.}(2012){Camero-Arranz}, {Finger},
  {Wilson-Hodge}, {Jenke}, {Steele}, {Coe}, {Gutierrez-Soto}, {Kretschmar},
  {Caballero}, {Yan}, {Rodr{\'{\i}}guez}, {Suso}, {Case}, {Cherry}, {Guiriec},
  \& {McBride}}]{ca12}
{Camero-Arranz}, A., {Finger}, M.~H., {Wilson-Hodge}, C.~A., {et~al.} 2012,
  \apj, 754, 20

\bibitem[{{Campana} {et~al.}(2002){Campana}, {Stella}, {Israel}, {Moretti},
  {Parmar}, \& {Orlandini}}]{campana02}
{Campana}, S., {Stella}, L., {Israel}, G.~L., {et~al.} 2002, ApJ, 580, 389

\bibitem[{{Chichkov} {et~al.}(1997){Chichkov}, {Sunyaev}, {Sazonov}, \&
  {Lund}}]{chichkov_disk}
{Chichkov}, M., {Sunyaev}, R., {Sazonov}, S., \& {Lund}, N. 1997, in ESA
  Special Publication, Vol. 382, The Transparent Universe, ed. C.~{Winkler},
  T.~J.-L. {Courvoisier}, \& P.~{Durouchoux}, 291

\bibitem[{{Churazov} {et~al.}(2001){Churazov}, {Gilfanov}, \&
  {Revnivtsev}}]{churazov}
{Churazov}, E., {Gilfanov}, M., \& {Revnivtsev}, M. 2001, \mnras, 321, 759

\bibitem[{Doroshenko(2011)}]{Doroshenko11}
Doroshenko, V. 2011, PhD thesis, Universit{\"a}t T{\"u}bingen

\bibitem[{{Doroshenko} {et~al.}(2010){Doroshenko}, {Santangelo}, {Suleimanov},
  {Kreykenbohm}, {Staubert}, {Ferrigno}, \& {Klochkov}}]{Doroshenko:2010p3661}
{Doroshenko}, V., {Santangelo}, A., {Suleimanov}, V., {et~al.} 2010, \aap, 515,
  A10+

\bibitem[{{Elsner} \& {Lamb}(1977)}]{elsner77}
{Elsner}, R.~F. \& {Lamb}, F.~K. 1977, \apj, 215, 897

\bibitem[{{Finger} {et~al.}(1996){Finger}, {Wilson}, \& {Harmon}}]{Finger96}
{Finger}, M.~H., {Wilson}, R.~B., \& {Harmon}, B.~A. 1996, \apj, 459, 288

\bibitem[{{Ghosh} \& {Lamb}(1979)}]{Ghosh:1979p1676}
{Ghosh}, P. \& {Lamb}, F.~K. 1979, \apj, 234, 296

\bibitem[{{Giangrande} {et~al.}(1980){Giangrande}, {Giovannelli}, {Bartolini},
  {Guarnieri}, \& {Piccioni}}]{Giangrande80}
{Giangrande}, A., {Giovannelli}, F., {Bartolini}, C., {Guarnieri}, A., \&
  {Piccioni}, A. 1980, \aaps, 40, 289

\bibitem[{{Giovannelli} {et~al.}(2007){Giovannelli}, {Bernabei}, {Rossi}, \&
  {Sabau-Graziati}}]{Giovannelli07}
{Giovannelli}, F., {Bernabei}, S., {Rossi}, C., \& {Sabau-Graziati}, L. 2007,
  \aap, 475, 651

\bibitem[{{Hill} {et~al.}(2007){Hill}, {Bird}, {Dean}, {McBride}, {Sguera},
  {Clark}, {Molina}, {Scaringi}, \& {Shaw}}]{hill07}
{Hill}, A.~B., {Bird}, A.~J., {Dean}, A.~J., {et~al.} 2007, \mnras, 381, 1275

\bibitem[{{Hoshino} \& {Takeshima}(1993)}]{fbreak0}
{Hoshino}, M. \& {Takeshima}, T. 1993, \apjl, 411, L79

\bibitem[{{Ikhsanov}(2001)}]{ikhsanov01a}
{Ikhsanov}, N.~R. 2001, A\&A, 367

\bibitem[{{Illarionov} \& {Sunyaev}(1975)}]{Illarionov:1975p2044}
{Illarionov}, A.~F. \& {Sunyaev}, R.~A. 1975, \aap, 39, 185

\bibitem[{{Lyubarskii}(1997)}]{lyubarskii_flicker}
{Lyubarskii}, Y.~E. 1997, \mnras, 292, 679

\bibitem[{{Motch} {et~al.}(1991){Motch}, {Stella}, {Janot-Pacheco}, \&
  {Mouchet}}]{Motch91}
{Motch}, C., {Stella}, L., {Janot-Pacheco}, E., \& {Mouchet}, M. 1991, \apj,
  369, 490

\bibitem[{{Mukherjee} \& {Paul}(2005)}]{Mukherjee05}
{Mukherjee}, U. \& {Paul}, B. 2005, \aap, 431, 667

\bibitem[{{Nagase} {et~al.}(1982){Nagase}, {Hayakawa}, {Kunieda}, {Makino},
  {Masai}, {Tawara}, {Inoue}, {Kawai}, {Koyama}, {Makishima}, {Matsuoka},
  {Murakami}, {Oda}, {Ogawara}, {Ohashi}, {Shibazaki}, {Tanaka}, {Miyamoto},
  {Tsunemi}, {Yamashita}, \& {Kondo}}]{Nagase:1982p2720}
{Nagase}, F., {Hayakawa}, S., {Kunieda}, H., {et~al.} 1982, \apj, 263, 814

\bibitem[{{Naik} {et~al.}(2008){Naik}, {Dotani}, {Terada}, {Nakajima},
  {Mihara}, {Suzuki}, {Makishima}, {Sudoh}, {Kitamoto}, {Nagase}, {Enoto}, \&
  {Takahashi}}]{Naik08}
{Naik}, S., {Dotani}, T., {Terada}, Y., {et~al.} 2008, \apj, 672, 516

\bibitem[{{Negueruela} {et~al.}(2000){Negueruela}, {Reig}, {Finger}, \&
  {Roche}}]{negu00}
{Negueruela}, I., {Reig}, P., {Finger}, M.~H., \& {Roche}, P. 2000, A\&A, 356

\bibitem[{{Orlandini} {et~al.}(2004){Orlandini}, {Bartolini}, {Campana}, {del
  Sordo}, {de Martino}, {Frontera}, {Guarnieri}, {Israel}, {Masetti},
  {Palazzi}, {Piccioni}, {Santangelo}, \& {Stella}}]{orlandini04}
{Orlandini}, M., {Bartolini}, C., {Campana}, S., {et~al.} 2004, Nuclear Physics
  B Proceedings Supplements, 132, 476

\bibitem[{{Perna} {et~al.}(2006){Perna}, {Bozzo}, \&
  {Stella}}]{Perna:2006p2027}
{Perna}, R., {Bozzo}, E., \& {Stella}, L. 2006, \apj, 639, 363

\bibitem[{{Postnov} {et~al.}(2008){Postnov}, {Staubert}, {Santangelo},
  {Klochkov}, {Kretschmar}, \& {Caballero}}]{Postnov:2008p348}
{Postnov}, K., {Staubert}, R., {Santangelo}, A., {et~al.} 2008, \aap, 480, L21

\bibitem[{{Reig}(2011)}]{Reig11}
{Reig}, P. 2011, \apss, 332, 1

\bibitem[{{Revnivtsev} {et~al.}(2009){Revnivtsev}, {Churazov}, {Postnov}, \&
  {Tsygankov}}]{revnivtsev}
{Revnivtsev}, M., {Churazov}, E., {Postnov}, K., \& {Tsygankov}, S. 2009, \aap,
  507

\bibitem[{{Reynolds} \& {Miller}(2010)}]{Reynolds10}
{Reynolds}, M.~T. \& {Miller}, J.~M. 2010, \apj, 723, 1799

\bibitem[{{Rosenberg} {et~al.}(1975){Rosenberg}, {Eyles}, {Skinner}, \&
  {Willmore}}]{Rosenberg75}
{Rosenberg}, F.~D., {Eyles}, C.~J., {Skinner}, G.~K., \& {Willmore}, A.~P.
  1975, \nat, 256, 628

\bibitem[{{Rothschild} {et~al.}(2013){Rothschild}, {Markowitz}, {Hemphill},
  {Caballero}, {Pottschmidt}, {K{\"u}hnel}, {Wilms}, {F{\"u}rst}, {Doroshenko},
  \& {Camero-Arranz}}]{Rothschild13}
{Rothschild}, R., {Markowitz}, A., {Hemphill}, P., {et~al.} 2013, \apj, 770, 19

\bibitem[{{Rutledge} {et~al.}(2007){Rutledge}, {Bildsten}, {Brown},
  {Chakrabarty}, {Pavlov}, \& {Zavlin}}]{rutledge07}
{Rutledge}, R.~E., {Bildsten}, L., {Brown}, E.~F., {et~al.} 2007, ApJ, 658

\bibitem[{{Steele} {et~al.}(1998){Steele}, {Negueruela}, {Coe}, \&
  {Roche}}]{Steele98}
{Steele}, I.~A., {Negueruela}, I., {Coe}, M.~J., \& {Roche}, P. 1998, \mnras,
  297, L5

\bibitem[{{Suleimanov} {et~al.}(2011){Suleimanov}, {Poutanen}, {Revnivtsev}, \&
  {Werner}}]{Suleimanov11}
{Suleimanov}, V., {Poutanen}, J., {Revnivtsev}, M., \& {Werner}, K. 2011, \apj,
  742, 122

\bibitem[{{Syunyaev} \& {Shakura}(1977)}]{Syunaev77}
{Syunyaev}, R.~A. \& {Shakura}, N.~I. 1977, Soviet Astronomy Letters, 3, 138

\bibitem[{{Titarchuk}(1994)}]{Titarchuk:1994p2324}
{Titarchuk}, L. 1994, \apj, 434, 570

\bibitem[{{Tsygankov} {et~al.}(2012){Tsygankov}, {Krivonos}, \&
  {Lutovinov}}]{Tsygankov2012}
{Tsygankov}, S.~S., {Krivonos}, R.~A., \& {Lutovinov}, A.~A. 2012, \mnras, 421,
  2407

\end{thebibliography}
\end{document}